\documentclass[english,aip,jcp,reprint]{revtex4-2}
\usepackage[T1]{fontenc}
\usepackage[utf8]{inputenc}
\usepackage{babel}
\usepackage{amsmath}
\usepackage{graphicx}
\usepackage{setspace}
\usepackage[pdfusetitle, bookmarks=true,  bookmarksnumbered=false, bookmarksopen=false, breaklinks=false, pdfborder={0 0 1}, backref=false, colorlinks=false]{hyperref}

\makeatletter

\usepackage[version=4]{mhchem}
\usepackage{notes2bib}
\ifdefined\showcaptionsetup
 \PassOptionsToPackage{caption=false}{subfig}
\fi
\usepackage{subfig}
 \usepackage{booktabs}
\bibnotesetup{
note-name = ,
use-sort-key = false
}

\usepackage{cleveref}
\AtBeginDocument{%
\let\ref\cref
}
\makeatother

\begin{document}
\title{Reaching precise proton affinities in non-Born--Oppenheimer calculations}
\author{Luukas Nikkanen}
\affiliation{Department of Chemistry, University of Helsinki, P.O. Box 55, FI-00014
University of Helsinki, Finland}
\author{Susi Lehtola}
\email{susi.lehtola@alumni.helsinki.fi}

\affiliation{Department of Chemistry, University of Helsinki, P.O. Box 55, FI-00014
University of Helsinki, Finland}
\begin{abstract}
An attractive way to model nuclear quantum effects is to describe select nuclei quantum mechanically at the same level as the electrons.
This non-Born--Oppenheimer (non-BO) method is known by many names including the nuclear-electronic orbital (NEO) and the multicomponent method.
Two basis sets are typically used for such calculations: a nuclear basis set and an electronic basis set.

In this work, we investigate the convergence of non-BO proton affinities (PAs) with respect to the protonic and electronic basis sets.
PAs are a sensitive measure of the proton and electron densities.
We demonstrate that most protonic basis sets are sufficient for non-BO density-functional calculations of PAs, resulting in convergence to within 0.1 kcal/mol of the complete protonic basis set limit. This indicates that the truncation error is dominated by the electronic basis, and that smaller protonic basis sets could be developed.

We show that non-BO calculations should use uncontracted electronic basis sets on the quantum protons.
The contraction coefficients in typical electronic basis sets have been derived for point nuclear charge distributions, and uncontracting the electronic basis set on the quantized proton leads to significantly faster convergence to the electronic basis set limit.
Uncontraction leads to results of at least one $\zeta$-level higher quality with negligible additional computational cost in multiple diffuse basis set families: Jensen's polarization consistent aug-pc-X basis sets, Dunning's correlation-consistent aug-cc-pVXZ basis sets, as well as the Karlsruhe def2-XZPD basis sets.
In specific, the aug-pc-3 electronic basis set already affords PAs converged beyond 0.1 kcal/mol when uncontracted on the quantum proton.
\end{abstract}
\maketitle
\newcommand*\ie{{\em i.e.}}
\newcommand*\eg{{\em e.g.}}
\newcommand*\etal{{\em et al.}}
\newcommand*\citeref[1]{ref.~\citenum{#1}}
\newcommand*\citerefs[1]{refs.~\citenum{#1}}
\newcommand*\citeit[1]{\citeauthor{#1}\cite{#1}}
\newcommand*\citeitcomma[1]{\citeauthor{#1},\cite{#1}}
\newcommand*\citeitperiod[1]{\citeauthor{#1}.\cite{#1}}

\newcommand*\Erkale{{\sc Erkale}}
\newcommand*\Bagel{{\sc Bagel}}
\newcommand*\FHIaims{{\sc FHI-aims}}
\newcommand*\LibXC{{\sc LibXC}}
\newcommand*\Orca{{\sc Orca}}
\newcommand*\PySCF{{\sc PySCF}}
\newcommand*\PsiFour{{\sc Psi4}}
\newcommand*\Turbomole{{\sc Turbomole}}

\section{Introduction \label{sec:Introduction}}

Nuclear quantum effects, such as zero-point energy, proton delocalization, and quantum tunneling play crucial roles in many aspects of chemistry, biology, and material science.\cite{McMahon2003_S_833, Kaestner2013_WIRCMS_158, Meisner2016_ACIE_5400, Schreiner2020_TC_980}
Among the various methods that have been developed for modeling quantum nuclear effects,\cite{Markland2018_NRC_109} the non-Born--Oppenheimer\cite{Born1927_APB_457} (non-BO) approach is a promising avenue for capturing these effects in a cost-efficient fashion, following the pioneering protonic structure calculations of \citeit{Thomas1969_PR_90}.
In this approach, protons and electrons are described on the same footing, and the Schr\"odinger equation is solved for these coupled particles.

However, as the masses and charges of the quantum particles are simply input parameters, the same methodology can also be used for calculations on, \emph{e.g.}, positrons\cite{Tachikawa2001_CPL_269} or muons\cite{Moncada2012_CPL_209} in atoms and molecules.
As a result of the breadth of possible applications the method is known under several names.
While historical precedence could thus argue also for the use of the ``protonic structure'' terminology, we opt to use the term ``non-BO'' due to its wide use across physics and chemistry literature, and since Born and Oppenheimer were surely among the first to consider the problem.
Alternatively, the method can also be called pre-Born--Oppenheimer.\cite{Matyus2019_MP_590}
Furthermore, since the calculations involve many kinds of particles, the term ``multi-component'' is also widely used; yet, we disfavor its use as it can be confused with the two- and four-component approaches of relativistic quantum mechanics.\cite{Saue2011_C_94}
Next, a number of authors have given alternative names to these calculations.
For example, the works in \citerefs{Tachikawa1998_CPL_437, Tachikawa1998_IJQC_491, Shigeta1998_IJQC_659, Gidopoulos1998_PRB_2146, Tachikawa1999_IJQC_497, Kreibich2001_PRL_2984,  Tachikawa2002_CPL_494, Webb2002_JCP_4106, Bochevarov2004_MP_111, Udagawa2006_JCP_244105, Nakai2007_IJQC_2849, Imamura2008_JCC_735, Gonzalez2008_IJQC_1742, Kreibich2008_PRA_22501, Ishimoto2009_IJQC_2677, Goli2011_TCA_235, Kylaenpaeae2012_PRA_52506, Ellis2015_JCTC_188, CassamChenai2017_TCA_52, Reyes2019_IJQC_25705, Muolo2020_JCP_204103, Pavosevic2020_CR_4222, HammesSchiffer2021_JCP_30901, Holzer2024_C_202400120} all represent non-BO calculations, even though some of these works examine other types of quantum particles than protons.
In specific, we must point out that the  nuclear-electronic orbital (NEO) method\cite{Webb2002_JCP_4106, Pavosevic2020_CR_4222, HammesSchiffer2021_JCP_30901} that has recently been popularized in chemistry is tantamount to the protonic structure method of \citeit{Thomas1969_PR_90}, or non-Born--Oppenheimer calculations in general.

The core of the non-BO method is to describe some of the nuclei quantum mechanically at the same level as the electrons.\cite{Thomas1969_PR_90}
Non-BO methods can be formulated with any of the standard model chemistries: for example, Hartree--Fock,\cite{Thomas1969_PR_90, Tachikawa1998_CPL_437, Tachikawa1998_IJQC_491, Webb2002_JCP_4106, Bochevarov2004_MP_111} density functional theory (DFT),\cite{Pak2007_JPCA_4522} configuration interaction
theory,\cite{Thomas1969_PR_90, Tachikawa2002_CPL_494, Webb2002_JCP_4106, Bochevarov2004_MP_111} perturbation theory,\cite{Bochevarov2004_MP_111, Nakai2003_JCP_1119, Swalina2005_CPL_394, Hoshino2006_JCP_194110} and coupled-cluster theory.\cite{Monkhorst1987_PRA_1544, Nakai2003_JCP_1119, Pavosevic2018_JCTC_338, Pavosevic2019_JCP_161102}
Development of novel non-BO methodologies is still an active area of research, as demonstrated by the recent implementations of local density fitting Hartree--Fock,\citep{Hasecke2023_JCTC_8223} local correlation MP2,\citep{Hasecke2024_JCTC_9928} Green's function approaches,\citep{Holzer2024_C_202400120} and double hybrid functionals,\cite{Hasecke2025_JCTC_11509} for example.

Typical non-BO models feature two sets of molecular orbitals: one for the electrons, and another for the protons (or the other type of quantum particle).
Typically, each of these is expanded in a basis set
\begin{align}
|\psi_{i}^\textrm{e}\rangle= & \sum_{\alpha=1}^{N_\text{bf}^\textrm{e}} C_{\alpha i}^\textrm{e}|\chi^\textrm{e}_{\alpha}\rangle\label{eq:basis-expansion-e}\\
|\psi_{I}^\textrm{p}\rangle= & \sum_{B=1}^{N_\text{bf}^\textrm{p}}C_{BI}^\textrm{p}|\chi_{B}^\textrm{p}\rangle\label{eq:basis-expansion-p}
\end{align}
where ${\bf C}^\textrm{e}$ and ${\bf C}^\textrm{p}$ are the electronic and protonic molecular orbital coefficients.
Note that while an electronic basis appears in \cref{eq:basis-expansion-e}, \cref{eq:basis-expansion-p} features a protonic basis set.

Even with the activity within the development of non-BO methodologies, it appears that there has been little work regarding protonic and electronic basis sets, or the study of basis set convergence in non-BO methods.

For the electrons it appears that most literature applications of the non-BO method in the NEO variant employ standard quantum chemical basis sets, such as Dunning's correlation consistent cc-pVXZ basis sets\citep{Dunning1989_JCP_1007} or the Karlsruhe def2 basis sets,\citep{Weigend2003_JCP_12753} as exemplified by the review of \citeit{Pavosevic2020_CR_4222} and a number of other recent publications.\cite{Yu2020_JCP_244123, Zhao2020_JCP_224111, Zhao2020_JPCL_4052, Yu2020_JPCL_10106, Pavosevic2021_JCTC_3252, Tao2021_ACR_4131, Tao2021_JCTC_5110, Schneider2021_JCP_54108, Pavosevic2021_JPCL_1631, Yu2022_JCP_114115, Pavosevic2022_JCP_74104, Liu2022_JPCA_7033, Li2022_JCTC_2774, Xu2022_JCP_224111, Feldmann2023_JCTC_856, Hasecke2023_JCTC_8223, Li2023_JCP_114118, Lambros2023_JPCL_2990, Dickinson2023_JPCL_6170, Chow2023_JPCL_9556, Liu2023_JCTC_6255}
Many works combine larger electronic basis sets centered on the quantum protons with smaller electronic basis sets centered on the classical nuclei, \emph{e.g.}, cc-pV5Z on the quantum protons and cc-pVDZ on other nuclei.\cite{Yang2018_JPCL_1765}

For the protons, early works employed various protonic basis sets: for example, even-tempered basis sets generated from the vibrational frequencies of the hydrogen molecule,\citep{Nakai2002_IJQC_511} double-$\zeta$ s, p, d nuclear (DZSDPN) basis sets,\citep{Webb2002_JCP_4106} or minimal $1s$ protonic basis sets.\cite{Ishimoto2006_JCP_144103, Ishimoto2008_JCP_164118}
Two types of protonic basis sets appear to be widely used in recent works.
The first are uncontracted even-tempered basis sets with the same exponents $\{\alpha_i=\alpha_0 \beta^{i-1}\}_{i=0}^{n-1}$ on the s, p and d shells, such as the 8s8p8d basis set of \citeit{Yang2017_JCP_114113} ($\alpha_0=2\sqrt{2}$, $\beta = \sqrt{2}$), and the closely related 2s2p2d, 4s4p4d, 6s6p6d, 8s8p8d, and 10s10p10d sets ($\alpha_0=4$, $\beta = 2$) recently studied by \citeit{Khan2025_JCC_70082}.
The second are the 10 protonic basis (PB) sets (PB4-D, PB4-F1, PB4-F2, PB5-D, PB5-F, PB5-G, PB6-D, PB6-F, PB6-G, and PB6-H) of \citeitcomma{Yu2020_JCP_244123} which are likewise uncontracted.

\begin{figure}
\begin{centering}
\includegraphics[width=\columnwidth]{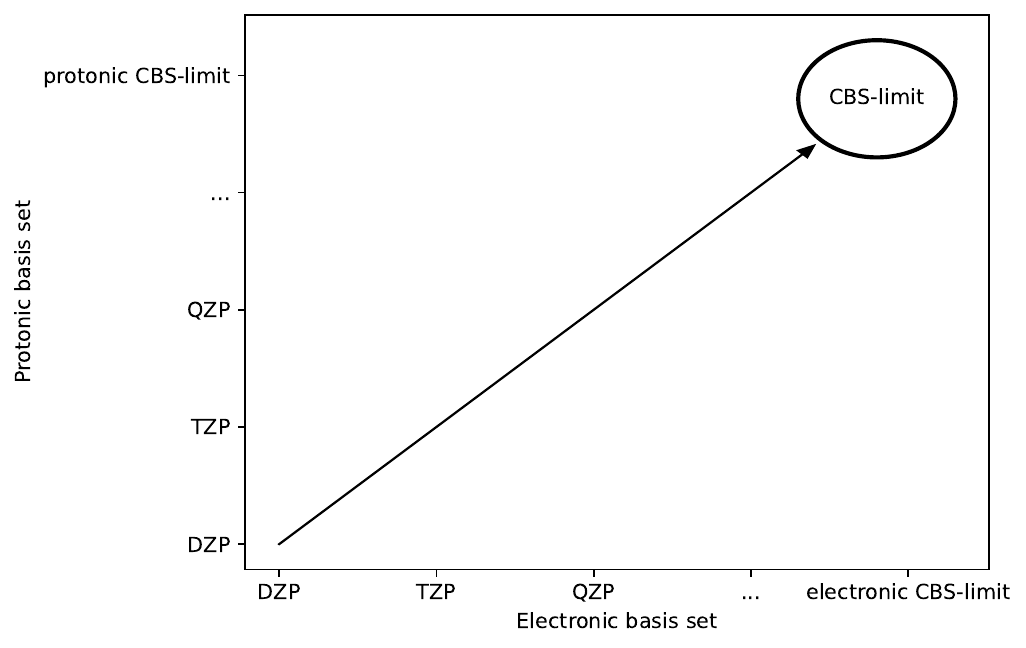}
\end{centering}
\caption{Pople diagram\cite{Pople1965_JCP_229} illustrating the CBS limit in non-BO calculations, which require going to the complete protonic and the complete electronic basis set.}
\label{fig:pople}
\end{figure}

Because non-BO methods involve two distinct types of basis sets, the overall complete basis set (CBS) limit is obtained at the simultaneous CBS limit of both types of basis sets, as illustrated by the Pople diagram\cite{Pople1965_JCP_229} in \cref{fig:pople}.
In this study, we investigate the basis set convergence of proton affinities (PAs) for a set of 13 molecules with non-BO-DFT.
PAs were chosen as the topic of this study, since they are a sensitive measure of proton and electron densities, which is also why they have been examined in many non-BO studies in the last decade.\cite{PedrazaGonzalez2016_PCCP_27185, PedrazaGonzalez2017_PCCP_25324, Brorsen2017_JPCL_3488, Brorsen2018_JCP_44110, Pavosevic2019_JCP_161102, Tao2019_JCP_124102, Rodas2019_JMM_316, Fajen2020_JCP_194107, Brorsen2020_JCTC_2379, Yu2020_JCP_244123, Pavosevic2020_JPCL_1578, Fajen2021_JCP_234108, Pavosevic2021_JPCL_1631, Pavosevic2022_JCP_74104, Fowler2022_JCTC_7298, Fetherolf2022_JPCL_5563, Goudy2025_JCP_224119, Hasecke2025_JCTC_11509, Khan2025_JCC_70082}
Our main focus is on the Jensen family of polarization-consistent basis sets\cite{Jensen2001_JCP_9113} optimized for density-functional theory (DFT),\cite{Hohenberg1964_PR_864, Kohn1965_PR_1133} but we also present results for the correlation consistent basis sets of \citeitcomma{Dunning1989_JCP_1007} as well as the Karlsruhe basis sets of \citeitperiod{Weigend2005_PCCP_305}
Herein, we show that uncontracting the electronic basis set of the quantum proton(s) in non-BO calculations significantly enhances the precision of the calculations, speeding up the basis set convergence in a remarkable fashion.
Uncontracted electronic basis sets have previously been considered in the non-BO context by \citeit{Nakai2003_JCP_1119}, \citeit{Nakai2005_JCP_164101}, and very recently by \citeit{Moncada2025_JCP_24110}, who note a slight improvement in the total energy estimates, but the practice does not appear to have become widespread.
We show that the uncontraction gives similar enhancement in proton affinity as the multicomponent-specific basis set by \citeitcomma{Samsonova2023_AO_5033} but does so at a negligible computational cost.

The layout of the manuscript is as follows:
In \cref{sec:Theory}, we summarize the theory behind the present calculations.
Then, we provide the computational details of the calculations of this work in \cref{sec:Computational-Details}.
We discuss the results of the calculations in \cref{sec:Results}, and finish with a brief summary and discussion in \cref{sec:summary}.
Hartree atomic units are used throughout the text, unless specified otherwise.

\section{Theory \label{sec:Theory}}

As already mentioned above, we employ the non-BO-DFT level of theory in this work.
Just like electronic DFT is justified by the Hohenberg--Kohn theorems,\cite{Hohenberg1964_PR_864} non-BO-DFT is in principle exact as shown by \citeitperiod{Capitani1982_JCP_568}
The non-BO-DFT energy functional is similar to that employed in electronic DFT (see \citeref{Lehtola2020_M_1218} for an overview); the main difference is that now in addition to the electron--electron exchange-correlation functional one also has a proton--proton exchange-correlation functional.
Typically in the NEO literature, one assumes the protons to be in a high-spin state and employs Hartree--Fock to avoid self-interaction errors,\cite{Pavosevic2020_CR_4222} and this is also what we do in this work.
Furthermore, one also has an electron--proton correlation functional to describe the strong interactions between the protonic and electronic parts of the total wave function; see \citerefs{Yang2017_JCP_114113}, \citenum{Brorsen2017_JPCL_3488}, and \citenum{Brorsen2018_JCP_44110} for commonly-used variants.

Following the standard linear combination of atomic orbitals (LCAO) approach, the minimization of the non-BO-DFT energy with respect to the electronic and protonic orbital coefficients in \cref{eq:basis-expansion-e,eq:basis-expansion-p} leads to the coupled eigenvalue problems
\begin{equation}
\begin{cases}
\mathbf{F}^\textrm{e}\mathbf{C}^\textrm{e}= & \mathbf{S}^\textrm{e}\mathbf{C}^\textrm{e}\mathbf{E}^\textrm{e}\\
\mathbf{F}^\textrm{p}\mathbf{C}^\textrm{p}= & \mathbf{S}^\textrm{p}\mathbf{C}^\textrm{p}\mathbf{E}^\textrm{p}
\end{cases}
\label{eq:roothaan}
\end{equation}
where $\mathbf{F}^\textrm{e}=\mathbf{F}^\textrm{e}(\mathbf{C}^\textrm{e},\mathbf{C}^\textrm{p})$
and $\mathbf{F}^\textrm{p}=\mathbf{F}^\textrm{p}(\mathbf{C}^\textrm{e},\mathbf{C}^\textrm{p})$
are the electronic and protonic Fock matrices that depend on both types of orbitals.
\Cref{eq:roothaan} presents the electronic spin-restricted case; the unrestricted case splits the electronic equation into a coupled problem for the spin-$\alpha$ and spin-$\beta$ electrons.

In conventional quantum chemistry with spherical coordinates centered at the nucleus, Gaussian basis functions of the type
\begin{align}
\langle {\bf r}|\chi_{i}\rangle= & N_{i} r^{l_i} \exp(-\alpha_i r^2) Y_{l_i}^{m_i}(\theta,\varphi)
\label{eq:prim-gto}
\end{align}
are used for the LCAO expansions.
In \ref{eq:prim-gto}, $N_{i}$ is a normalization constant, and $Y_{l}^{m}(\theta,\varphi)$ is a spherical harmonic.
The exponents are defined per angular momentum, and a full shell of functions with $m=-l,\dots, l$ are added for each exponent.
Hundreds of Gaussian-type orbital basis sets have been developed for electrons during the last several decades\cite{Davidson1986_CR_681, Hill2013_IJQC_21, Jensen2013_WIRCMS_273} and most of them are available on the Basis Set Exchange.\cite{Pritchard2019_JCIM_4814}

While all protonic basis sets used so far appear to follow the uncontracted form of \cref{eq:prim-gto}, most electronic basis sets employ contractions to reduce the necessary total number of basis functions in typical calculations.
In the Born--Oppenheimer case, a proton at ${\bf R}$ gives rise to the following potential for the electrons
\begin{equation}
  \label{eq:epcoulomb}
  V({\bf r}) = -\frac 1 {|{\bf r}-{\bf R}|},
\end{equation}
which diverges $V({\bf r}) \to -\infty$ when approaching the nucleus, ${\bf r} \to {\bf R}$.
Hence, at close range the potential is largely independent of the chemical environment.
Molecular orbitals thus tend to behave similarly to atomic orbitals close to the nuclei (core orbitals are insensitive to the chemical environment and do not participate significantly to chemical bonding).
This motivates contracting the basis set: tight functions always appear in a similar combination regardless of the chemical environment.
Contracting the basis set induces a so-called contraction error; yet, the contractions are typically chosen in a way to make the contraction error negligible with respect to the overall basis set truncation error (BSTE) of the finite basis set in calculations of chemically relevant relative energies.

However, as we will show in \cref{sec:Results}, the quantum nuclei in non-BO calculations are no longer modeled as point charges, and the electron-nuclear Coulomb potential will behave differently close to the center of the nuclear charge distribution.
The potential is now given by
\begin{equation}
  \label{eq:epcoulomb-neo}
  V({\bf r}) = -\int \frac {n^p({\bf r}')} {|{\bf r}-{\bf r}'|} {\rm d}^3r',
\end{equation}
where the electrons see the charge of a delocalized proton.
The proton density $n^p$ in \cref{eq:epcoulomb-neo} is no longer a static point charge distribution.
Instead, it is delocalized over a finite volume, and the potential of \cref{eq:epcoulomb-neo} is finite everywhere.

Moreover, the proton density $n^p$ is determined dynamically for each system, and is likely not transferable across systems.
Thus, we have reason to believe that the electronic basis sets for non-BO calculations need to carry extra flexibility in regions around the quantized nuclei.
Note that while the mean-field orbitals arising from \cref{eq:roothaan,eq:epcoulomb} no longer carry a cusp, Kato's cusp conditions\cite{Kato1957_CPAM_151} will still be valid for the exact solution, in that the exact wave function can be shown to have a special limit when the protonic and electronic coordinates coalesce, analogously to the case of the electron-electron cusp of exact theory.

Established experience with nuclear spin-spin-coupling\cite{Helgaker1998_TCA_175} and x-ray calculations\cite{Besley2009_JCP_124308} demonstrate that basis sets should not be contracted when studying properties that are sensitive to the near-core region.
Now, as non-BO calculations do not feature point nuclei, we check whether uncontracting the electronic basis set on the quantum protons (as in references \citenum{Nakai2003_JCP_1119}, \citenum{Nakai2005_JCP_164101}, and \citenum{Moncada2025_JCP_24110}) will also result in better basis set convergence in non-BO calculations.

For clarity, we also note here that the delocalization of the proton in non-BO calculations is unrelated to the physically known finite size of the proton, exhibiting a radius smaller than 1 fm.\cite{Carlson2015_PPNP_59, Hammer2020_SB_257}
The size of the proton itself is much smaller than the delocalization of the proton's wave function observed in typical systems, as we will demonstrate in \cref{sec:Results}.
The best analogy here is to electrons, which {\em are} thought to be point particles, but which still feature a delocalized probability density in systems.

Uncontracting the electronic basis set eliminates its contraction error and allows more freedom in the electrons' behavior close to the center of the nuclear distribution, greatly enhancing the electronic basis set convergence of the non-BO calculation.
In contrast, conventional electronic basis sets exhibit significant contraction errors in non-BO calculations, as is revealed by the comparison between the errors obtained with contracted and uncontracted electronic basis sets on the quantum protons.

\section{Computational Details \label{sec:Computational-Details}}

As already mentioned in \cref{sec:Introduction}, we study the basis set convergence of the PAs of 13 molecules: \ce{CN-}, \ce{NO2-}, \ce{NH3}, \ce{HCOO-}, \ce{H2O}, \ce{OH-}, \ce{H2S}, \ce{SH-}, \ce{CO}, \ce{N2}, \ce{CO2}, \ce{CH2O}, and \ce{2F-}.
In each protonated system, the most acidic proton is described quantum mechanically, while the Born--Oppenheimer approximation\citep{Born1927_APB_457} is assumed for the other nuclei.\cite{DiazTinoco2013_JCP_194108}
The PA for the most acidic hydrogen in each molecule is calculated as
\begin{equation}
\text{PA(A)}=E_{\ce{A}}-E_{\ce{AH+}}+\frac{5}{2}RT\label{eq:proton_affinity}
\end{equation}
where $E_{\ce{A}}$ is the Born--Oppenheimer energy for the deprotonated species, $E_{\ce{AH+}}$ is the non-BO energy of the protonated species, $R$ is the universal gas constant, and $T$ is the temperature that was set to 293.15 K.

All calculations were performed using NEO functionality in Q-Chem 6.1\citep{Epifanovsky2021_JCP_84801} using the B3LYP electronic exchange-correlation functional.\citep{Stephens1994_JPC_11623}
All of the basis sets were taken from the Basis Set Exchange\cite{Pritchard2019_JCIM_4814} via its Python interface, which was also used to form the uncontracted basis sets.
Both the electronic and protonic basis sets were used in spherical form, using the Q-Chem input keywords \texttt{PURECART 1111} and \texttt{NEO\_PURECART 1111}, respectively.
The geometries for the molecules and their protonated variants were optimized using conventional DFT for each of the electronic basis sets using standard settings in Q-Chem.
As the geometry optimizations did not involve non-BO calculations, they employed the basis sets in standard, contracted form.
The mixed-basis calculations of \ref{sec:mixedbasis} used geometries obtained from DFT calculations using the smaller (environment) basis set.
The employed optimized geometries are available in the supporting information (SI).

The optimized geometries were then used to carry out the non-BO-DFT calculations employing the epc17-2 electron--proton correlation functional\citep{Brorsen2017_JPCL_3488} and a (150,974) Euler--McLaurin--Lebedev quadrature grid,\cite{Murray1993_MP_997, Lebedev1976_UCMMP_10} which we found to yield suitably converged total energies.

As was also already mentioned in \cref{sec:Introduction}, we consider the following electronic basis set families: Dunning's correlation consistent (cc) family,\citep{Dunning1989_JCP_1007} Jensen's polarization consistent (pc) family,\citep{Jensen2001_JCP_9113} as well as the Karlsruhe def2 family.\citep{Weigend2005_PCCP_305}
Because six of the molecules considered herein are anions in their deprotonated state, diffuse functions were employed for all three families.\cite{Kendall1992_JCP_6796, Jensen2002_JCP_9234, Rappoport2010_JCP_134105}
For the calculations with the cc basis sets, we also consider the non-BO specific augmentations for the electronic basis for multicomponent (mc) calculations recently proposed by \citeitperiod{Samsonova2023_AO_5033}

We consider the 10 PB protonic basis sets (PB4-D, PB4-F1, PB4-F2, PB5-D, PB5-F, PB5-G,
PB6-D, PB6-F, PB6-G, and PB6-H) of \citeitcomma{Yu2020_JCP_244123} as well as the even-tempered 8s8p8d basis set proposed by \citeitperiod{Yang2017_JCP_114113}
We perform additional calculations with the analogous 8s, 8s8p, and 8s8p8d8f even-tempered protonic basis sets with the same universal exponents of the 8s8p8d basis set.
We also check the convergence of the even-tempered protonic basis sets with respect to steeper functions with the 10s10p10d10f basis set used for density fitting in the literature.\cite{Xu2022_JCP_224111}
We further supplement the study with the 2s2p2d, 4s4p4d, 6s6p6d, and 8s8p8d even-tempered sets of \citeitperiod{Khan2025_JCC_70082}
The obtained total energies for all of the above described calculations are also available in the SI.

\section{Results \label{sec:Results}}

The usual goal in quantum chemistry is to reach chemical accuracy, often considered to be within 1 kcal/mol of the exact theoretical or experimental result.\cite{Pople1999_RMP_1267}
However, it is important to distinguish here between the two possible sources of error: the basis set truncation error (related to \emph{precision}), and the error inherent in the method employed in the calculations (related to \emph{accuracy}), as differences between computational results and either experiment or accurate theoretical reference values are always a mixture of the two effects.
For this reason, it is important to be wary of Pauling points:\cite{Loewdin1986_IJQC_19}
Combinations of incomplete basis sets and inaccurate methods may lead to fortuitous error cancellation and a serendipitously small total error.
Yet, this small error is typically completely untransferable to the modeling of other systems.

Key to the basis set truncation error is that it can be made negligibly small by using a sufficiently large basis set.
We show in this section that the PAs can be converged to 0.1 kcal/mol in non-BO-DFT with respect to the electronic and protonic basis sets.
Access to such converged PAs then allow comparison to experiment, elucidating the error inherent in the employed density functional, and also enables the training and learning of more accurate density functionals for non-BO, for example.

Since the changes in the electronic structure induced by the quantum protons are likely localized to the close proximity of the quantum protons themselves, we will only study uncontractions of the electronic basis sets on the quantum protons, and use contracted basis functions on all the other nuclei.
We denote this scheme by the uncHq- prefix.

\subsection{Why Does Uncontraction Help? \label{sec:unc}}

\begin{figure*}
\subfloat[Electron density]{
    \centering{}
    \includegraphics[width=0.5\linewidth]{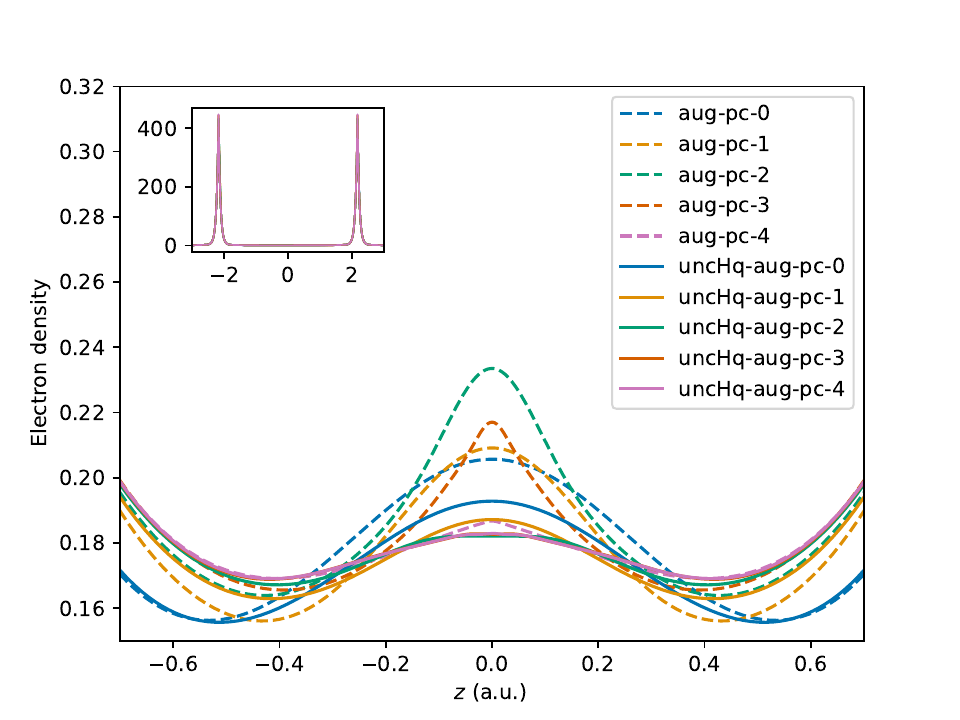}
    }
\subfloat[Proton density]{
    \centering{}
    \includegraphics[width=0.5\linewidth]{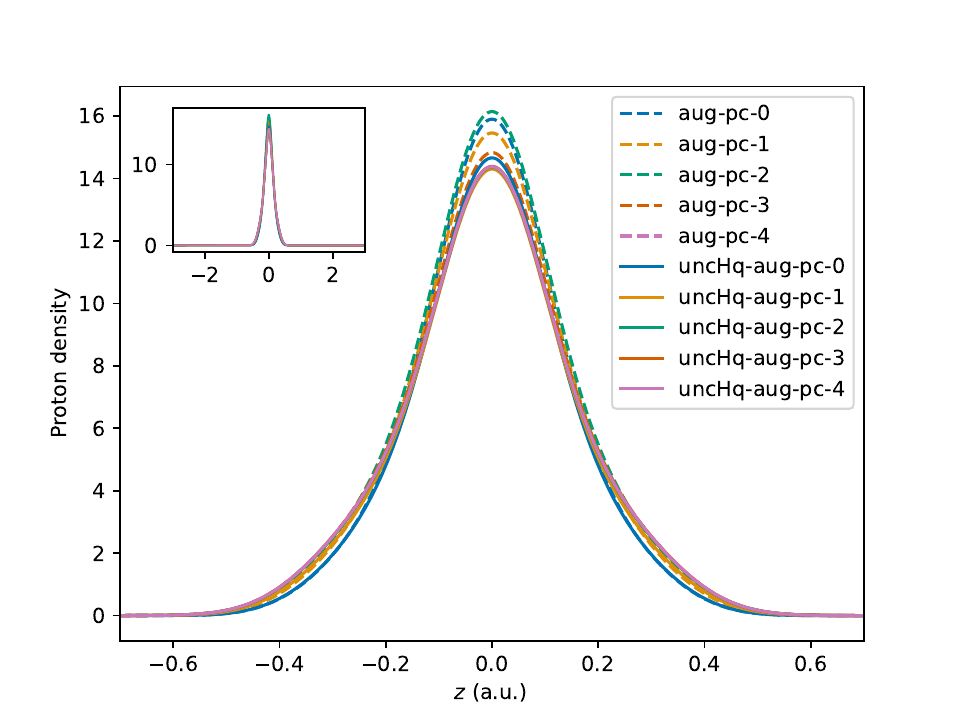}
    }

\caption{Electron (a) and proton densities (b) along the bond of
  \ce{FHF-} molecule for various electronic basis sets, with a fixed
  FH distance of 1.149 \AA. The B3LYP electronic density functional
  was employed with the epc17-2 electron-proton correlation functional
  and the PB4-F1 protonic basis set.}
\label{fig:protondensity}
\end{figure*}

To begin, we demonstrate the validity of the discussion in \cref{sec:Theory} by showing that uncontracting the electronic basis on the quantum hydrogen leads to significant improvement in the proton and electron densities.
For this demonstration, we choose the \ce{FHF-} molecule with a fixed FH distance 1.149 \AA, and plot the electron and proton densities along the bond.
For this part of the study, we fix the protonic basis set to PB4-F1, and start the analysis with Jensen's electronic aug-pc-$n$ basis sets which are optimized for DFT calculations.
These results are shown in \cref{fig:protondensity}.
The data show large errors in the electron and proton density in the vicinity of the delocalized quantum proton when the electronic basis is used in its original contracted form, while uncontracting the electronic basis leads to qualitatively correct densities already with the smallest considered electronic basis set.
The observed errors in the electron and proton densities with the contracted basis sets arise from the lack of flexibility of the electronic basis close the basis set's origin.
Note especially that the proton density changes even though the same protonic basis set is employed in all of the calculations.
The length scale of the proton delocalization in \cref{fig:protondensity} is also noteworthy; it is roughly five orders of magnitude greater than the proton's charge radius $r_p = 0.8406$ fm.\cite{Maisenbacher2026_N_845}

\begin{figure}
    \centering
    \includegraphics[width=\linewidth]{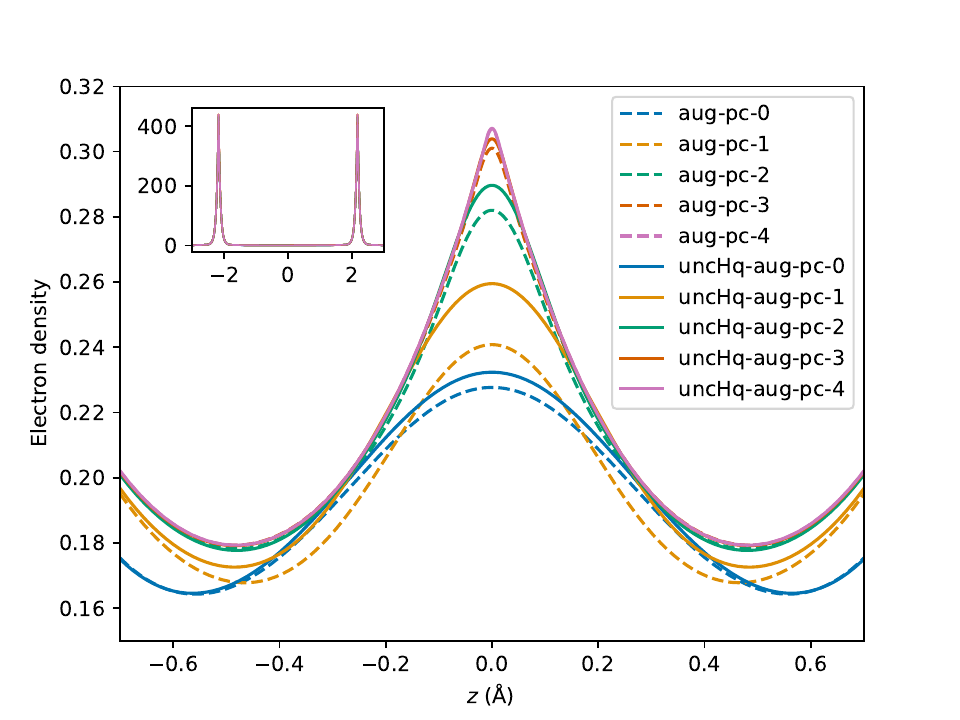}  
    \caption{Electron density along the bond of the \ce{FHF-} molecule with a fixed FH distance of 1.149 \AA. The B3LYP electronic density functional was employed with the BO approximation. }
    \label{fig:bo-h}
\end{figure}

The behavior observed in \cref{fig:protondensity} (whose physical background was discussed in \cref{sec:Theory}) stands in stark contrast with the classical case shown in \cref{fig:bo-h}, where only small differences are observed in the results obtained with the contracted and uncontracted basis sets.
Of course, the reason for this behavior is that the BO calculations converge to a cusp at the hydrogen at the CBS limit; this cusp is flattened in \cref{fig:bo-h} due to the use of finite Gaussian basis sets.

\subsection{Reaching the electronic basis set limit \label{sec:electroniclimit}}

\begin{figure}
\begin{centering}
\includegraphics[width=1\linewidth]{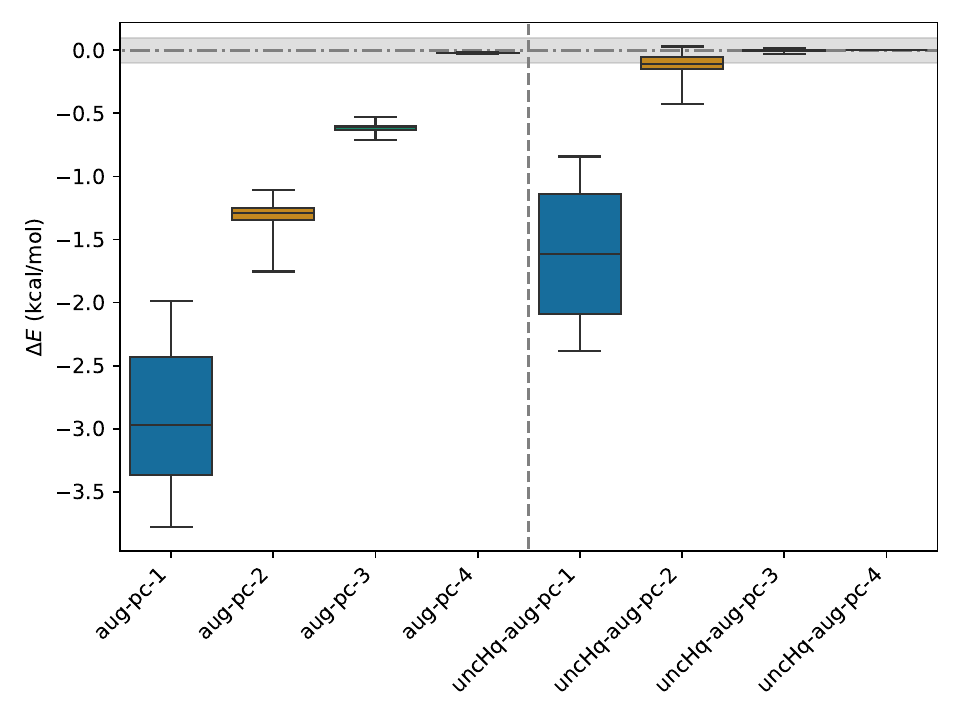}
\par\end{centering}
\centering{}
\caption{Box plot representing the BSTEs of PAs obtained with the aug-pc-$n$ family of electronic basis sets and the PB4-F1 protonic basis set.
The reference is the largest basis set studied, i.e., the uncHq-aug-pc-4/PB4-F1 values.
The box extends from the first quartile to the third quartile, and the line represents the median.
The whiskers show the full range of values.
The shaded area shows the range of our target precision of $\pm0.1$ kcal/mol.
}
\label{fig:pc}
\end{figure}

Although the CBS limit must be reached simultaneously for the protons and the electrons, we begin with the electronic problem.
As in \cref{sec:unc}, we keep the protonic basis set fixed to PB4-F1, and approach the electronic CBS limit with the aug-pc-$n$ basis sets.
\Cref{fig:pc} shows the errors in calculated PAs relative to uncHq-aug-pc-4 values for both the contracted aug-pc-$n$ basis sets and the partially uncontracted uncHq-aug-pc-$n$ basis sets, with the cardinal number $n$ of the basis sets varying from 1 to 4.
The reference values are shown in \cref{tab:pas}.

\begin{table}
\begin{tabular}{lrr}
\toprule
Molecule  & PB4-F1 & PB6-H \\
\midrule
\ce{CN-}   & 349.77  & 349.78       \\
\ce{NO2-}  & 339.28  & 339.29       \\
\ce{NH3}   & 205.58  & 205.58       \\
\ce{HCOO-} & 344.23  & 344.24       \\
\ce{OH-}   & 389.18  & 389.19       \\
\ce{SH-}   & 350.08  & 350.08       \\
\ce{H2O}   & 167.24  & 167.25       \\
\ce{H2S}   & 170.73  & 170.74       \\
\ce{CO}    & 142.83  & 142.83       \\
\ce{N2}    & 119.92  & 119.92       \\
\ce{CO2}   & 130.87  & 130.89       \\
\ce{CH2O}  & 173.03  & 173.04       \\
\ce{2F-}   & 416.69  & 416.66       \\
\bottomrule
\end{tabular}
\caption{B3LYP/epc17-2 PAs of the studied molecules. The data were obtained with the uncHq-aug-pc-4 electronic basis set and the PB4-F1 or PB6-H protonic basis sets. All values are in kcal/mol.}
\label{tab:pas}
\end{table}

Analysis of the results shows that the fully contracted polarization consistent basis sets lead to slow convergence of the PAs with respect to the cardinal number $n$ of the basis set.
The quadruple-$\zeta$ aug-pc-3 basis set is necessary to achieve a estimated basis set truncation error within chemical accuracy, and our final convergence criterion of 0.1 kcal/mol is only met at the quintuple-$\zeta$ level of aug-pc-4.
This quantitative finding is in line with the qualitative one from \cref{fig:protondensity}.

Uncontracting the electronic basis set on the quantum nuclei in the uncHq- basis sets leads to remarkable improvement in the speed of convergence: already the triple-$\zeta$ basis set is converged to within chemical accuracy and the quadruple-$\zeta$ has converged to the basis set limit with maximal errors of 0.42 kcal/mol and 0.03 kcal/mol respectively.
Conventional electronic basis sets in their contracted form thus are suboptimal when the proton is allowed to delocalize as a quantum particle.
This finding is again in agreement with \cref{fig:protondensity}.

Importantly, these improvements in precision are accompanied with negligible increase in computational cost.
Typically, only some of the hydrogen atoms are modeled as quantum particles, and uncontracting the electronic basis on a quantum hydrogen introduces a mere 2, 3, 4, and 4 basis functions in the double- to quintuple-$\zeta$ aug-pc-1, aug-pc-2, aug-pc-3, and aug-pc-4 basis sets, respectively, while the unmodified basis sets have a total of 9, 23, 50, and 88 basis functions per hydrogen atom.

\begin{figure}
\begin{centering}
\includegraphics[width=1\linewidth]{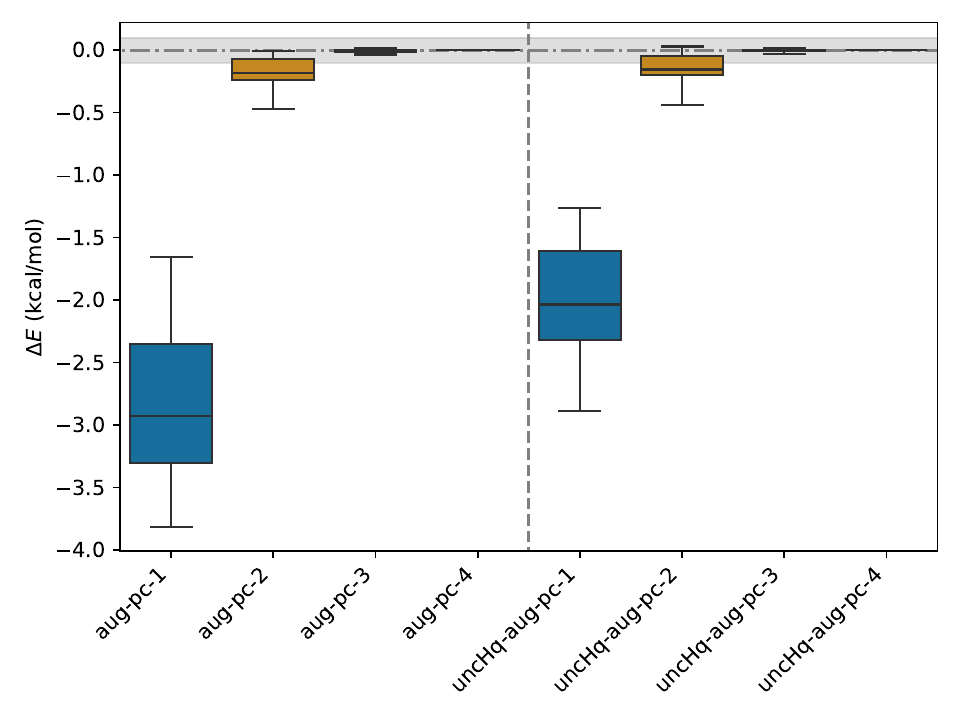}
\par\end{centering}
\centering{}
\caption{BSTEs of PAs obtained with the aug-pc-$n$ family using conventional BO calculations. All data are relative to the uncHq-aug-pc-4 values. Even though these calculations feature no quantum protons, the electronic basis set on the most acidic hydrogen is uncontracted in analogy to the results of \cref{fig:pc}. The distributions for the quadruple- and quintuple-$\zeta$ basis sets (aug-pc-3, uncHq-aug-pc-3, aug-pc-4) are sharply focused around zero. The plot and box parameters are the same as in \cref{fig:pc}}
\label{fig:pc_classical}
\end{figure}

With the exclusion of the double-$\zeta$ aug-pc-1 results, the data in \cref{fig:pc} are in stark contrast with the case with only classical nuclei shown in \ref{fig:pc_classical}, where the difference in results between contracted and uncontracted basis sets are negligible; after all, the basis sets---including their contraction patterns---are designed to exhibit negligible contraction errors for standard calculations, like the determination of proton affinities.
A similar result was observed above in \cref{fig:bo-h}.

\subsection{Reaching the protonic basis set limit \label{sec:protoniclimit}}

\begin{figure}
\begin{centering}
\includegraphics[width=\linewidth]{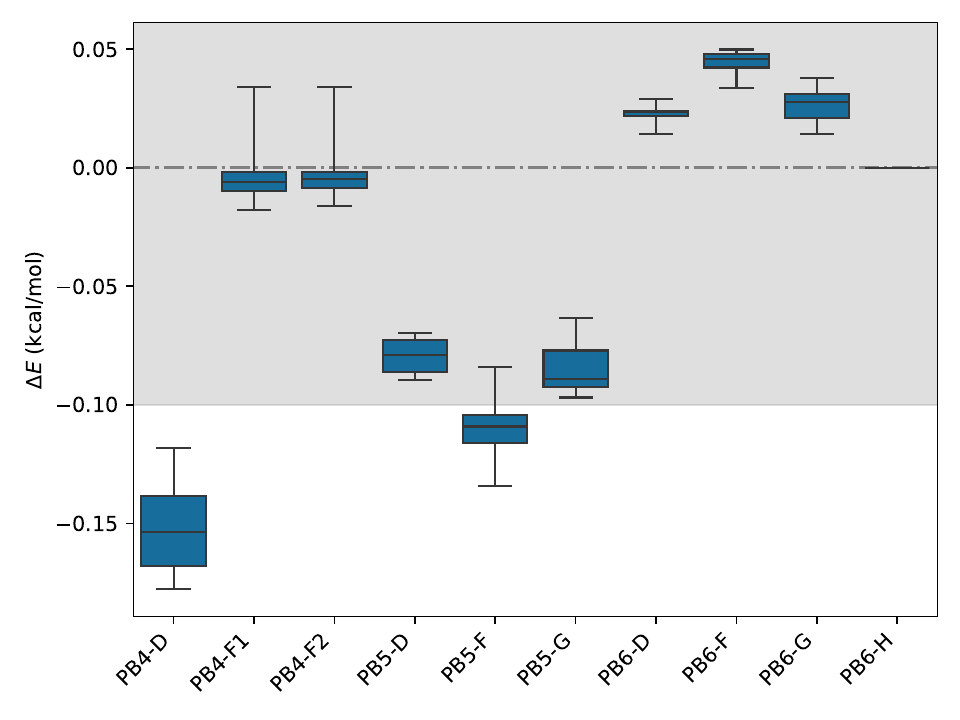}
\end{centering}
\caption{BSTEs of PAs obtained with the protonic PB set family and the electronic uncHq-aug-pc-4 basis set. All data are reported relative to uncHq-aug-pc-4/PB6-H values. Observe the small scale of the $y$-axis. The plot and box parameters are the same as in \cref{fig:pc}}
\label{fig:PB}
\end{figure}

Having established that the uncHq-aug-pc-4 results are converged to the electronic basis set limit, we investigate the convergence behavior of the protonic basis sets.
We begin by examining the ten PB sets of \citeitcomma{Yu2020_JCP_244123} using the value obtained with the largest of the PB sets (PB6-H) as reference; the PB6-H values are also shown in \cref{tab:pas}.
The results for the various PB sets are shown in \cref{fig:PB}.

The PB sets do not represent a systematic hierarchy for PAs.
The PB sets have been stochastically optimized for a linear combination of three properties: non-BO coupled-cluster energy, non-BO-DFT proton densities \emph{vs.} proton densities from the BO Fourier grid Hamiltonian (FGH) method, and non-BO time-dependent Hartree--Fock (TD-HF) proton vibrational excitation energies \emph{vs.} BO FGH vibrational energies.
While the first of these terms is fine, the second and third terms exhibit an inconsistent use of methodologies which is bound to lead to convergence to Pauling points: non-BO calculations cannot be expected to agree with BO calculations, especially when they employ further inexact approximations.

Regardless, PB4-F1 and PB4-F2 give results that appear roughly equally well converged, whereas the PB5 basis sets that are larger than the analogous PB4 sets clearly have a poorer level of precision and show non-systematic behavior of the truncation error going from the smallest PB5-D basis set to the larger PB5-F and PB5-G basis sets.
The PB6 sets again exhibit better agreement with the PB6-H values, similarly to the PB4-F1 and PB4-F2 basis sets, but the convergence is again non-systematic.

These findings can be rationalized by an in-depth examination of the PB sets.
A peculiar feature of them is that despite the similar naming, the s, p, and d exponents in PB4-D differ from those in PB4-F1 and PB4-F2, even though the latter two only differ by PB4-F2 having an additional f function with exponent 20.985.
Similarly, the PB5 and PB6 basis sets are all distinct, which explains the fluctuations observed in the data.

Despite these fluctuations, we observe already from the PB data that the differences between the PAs predicted by the various PB sets are small, often satisfying our aimed convergence criterion of 0.1 kcal/mol.

\begin{figure}
\begin{centering}
\includegraphics[width=\linewidth]{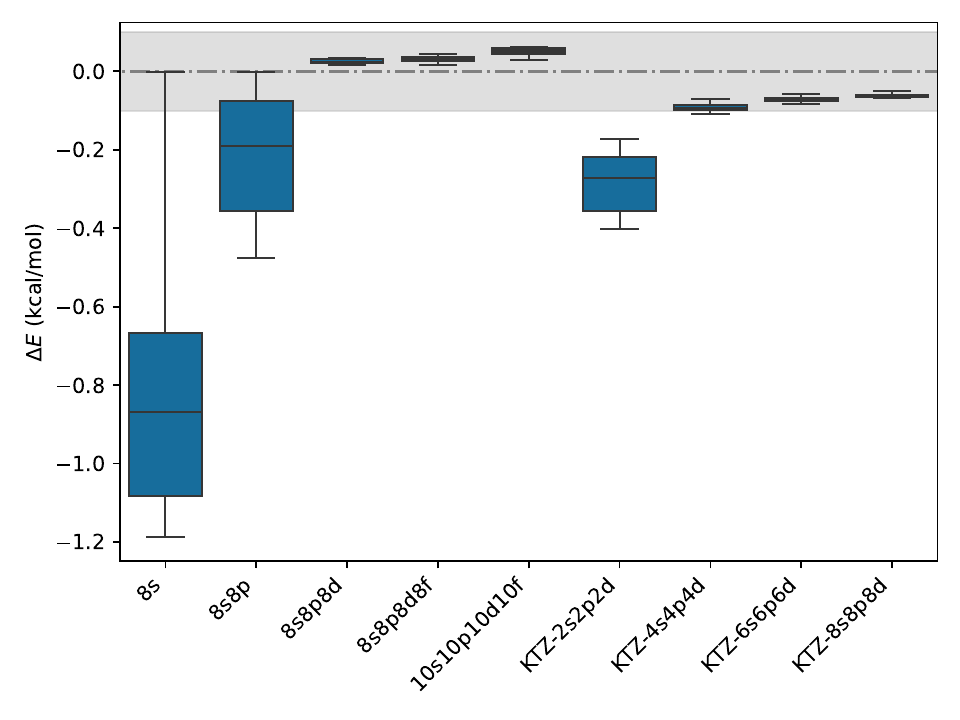}
\end{centering}
\caption{BSTEs of PAs obtained with the even-tempered protonic basis sets and the electronic uncHq-aug-pc-4 basis set. All data are reported relative to uncHq-aug-pc-4/PB6-H values. KTZ refers to the basis sets of \citeitperiod{Khan2025_JCC_70082} The plot and box parameters are the same as in \cref{fig:pc}}
\label{fig:epc}
\end{figure}

For an independent confirmation, we repeat the analysis with the $\{2\sqrt{2}^i\}_{i=1}^n$ even-tempered 8s, 8s8p, 8s8p8d, and 8s8p8d8f ($n=8$), and 10s10p10d10f ($n=10$) protonic basis sets of \citeitcomma{Yang2017_JCP_114113} as well as the $\{2^{i+1}\}_{i=1}^n$ even-tempered 2s2p2d, 4s4p4d, 6s6p6d, and 8s8p8d sets of \citeit{Khan2025_JCC_70082}.
The data for these calculations are shown in \cref{fig:epc}.
The small difference between the 8s8p8d8f and 10s10p10d10f results and the PB4-F1 data again suggests that the protonic basis set limit has been reached.
Given that the 8s8p8d and PB6-D basis sets show relatively small errors, we believe the exponents in the PB4-D, PB5-D, PB5-F, and PB5-G basis sets to be suboptimal.
In agreement with \citeit{Khan2025_JCC_70082}, we find that the even-tempered 4s4p4d set with exponents 4, 8, 16, and 32 is adequate to reproduce PAs at the non-BO-DFT level; after all, the minimal protonic 1s basis of \citeit{Lehtola2025_JPCA_5651} is snugly included in this range of exponents.
The comparison of the 8s8p8d sets of \citeit{Yang2017_JCP_114113} and \citeit{Khan2025_JCC_70082} shows that the former leads to more tightly bound protons, and thus exhibits a better range of exponents.
As neither of these basis sets appears to have been tightly optimized, it appears that further improvements in precision and cost-efficiency are possible by the development of dedicated basis sets.

\subsection{Cross-assessment, transferability of results and lack of functional dependence \label{sec:fucdep}}

Finally, we demonstrate that the simultaneous electronic and protonic CBS limit has been reached, and that our results are transferable across functionals.
BO-DFT calculations are well-known to exhibit exponential convergence to the basis set limit, and the same appears to be true of non-BO-DFT calculations as well.
For this part of the study, we consider the Perdew--Wang local-density approximation\cite{Bloch1929_ZP_545, Dirac1930_MPCPS_376, Perdew1992_PRB_13244} (PW92), the hybrid\cite{Adamo1999_JCP_6158, Ernzerhof1999_JCP_5029} of the Perdew--Burke--Ernzerhof generalized-gradient approximation functional\cite{Perdew1996_PRL_3865, Perdew1997_PRL_1396} (PBE0), as well as the r$^2$SCAN meta-generalized gradient approximation functional\cite{Furness2020_JPCL_8208, Furness2020_JPCL_9248} in addition to B3LYP.
Together, these functionals span the first four rungs of Jacob's ladder.\cite{Perdew2001_ACP_1}
Furthermore, we also consider Hartree--Fock (HF) as the fifth choice; alike DFT, HF also converges exponentially to the basis set limit.
We consider two choices for the electron-proton correlation functional: omitting it, and the epc17-2 functional,\cite{Brorsen2017_JPCL_3488} which remains the only electron-proton correlation functional used in practice.

\begin{figure*}
    \centering
    \includegraphics[width=.7\textwidth]{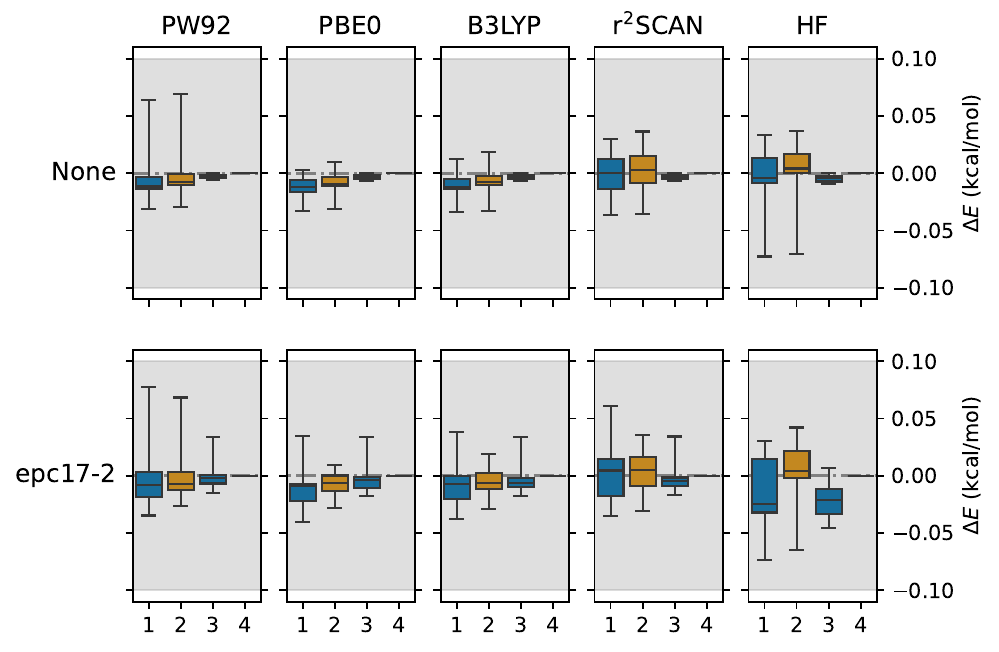}  
    \caption{Basis set truncation errors relative to the largest basis (uncHq-aug-pc-4+PB6-H) with all combinations of 5 electronic methods (PW92, PBE0, B3LYP, r$^2$SCAN, and HF) with two electron-proton correlation functionals (none and epc17-2). The basis set combinations on the $x$ axis are as follows: uncHq-aug-pc-3+PB4-F1 (1), uncHq-aug-pc-3+PB6-H (2), uncHq-aug-pc-4+PB4-F1 (3), and uncHq-aug-pc-4+PB6-H (4).}
    \label{fig:func_dependency}
\end{figure*}

For all of these combinations, we show that the basis set truncation error with the uncHq-aug-pc-3 or the uncHq-aug-pc-4 electronic basis set and the PB4-F1 or PB6-H protonic basis set is smaller than 0.1 kcal/mol compared to the corresponding uncHq-aug-pc-4/PB6-H result.
The resulting data illustrated in \cref{fig:func_dependency} emphatically show that this is indeed the case.
As all differences are completely negligible and well inside our aimed basis set convergence criterion of 0.1 kcal/mol, this proves that we have reached the CBS limit for the non-BO-DFT PAs.
The corresponding CBS limit PAs are given in \cref{tab:other-pas}, augmenting the data in \cref{tab:pas}.

\begin{table*}
\setlength{\tabcolsep}{3.7pt} 
\begin{tabular}{lccccccccccc}
\toprule
 & \multicolumn{2}{c}{PW92} & \multicolumn{2}{c}{PBE0} & B3LYP & \multicolumn{2}{c}{r$^2$SCAN} & \multicolumn{2}{c}{HF} \\
 \cmidrule(lr){2-3}\cmidrule(lr){4-5}\cmidrule(lr){6-6}\cmidrule(lr){7-8}\cmidrule(lr){9-10}
 & none & epc17-2 & none & epc17-2 & none & none & epc17-2 & none & epc17-2 \\
\midrule
\ce{CN-}   & 326.06 & 344.22 & 333.25 & 351.36 & 331.78 & 332.66 & 350.75 & 332.61 & 350.76 \\
\ce{NO2-}  & 313.41 & 331.85 & 322.60 & 341.00 & 320.98 & 323.40 & 341.78 & 328.76 & 347.29 \\
\ce{NH3}   & 183.67 & 201.90 & 189.39 & 207.57 & 187.49 & 189.20 & 207.37 & 192.29 & 210.54 \\
\ce{HCOO-} & 318.92 & 337.38 & 327.93 & 346.36 & 325.90 & 327.04 & 345.46 & 334.51 & 353.09 \\
\ce{OH-}   & 363.82 & 382.37 & 374.45 & 392.94 & 370.81 & 372.94 & 391.42 & 383.04 & 401.68 \\
\ce{SH-}   & 325.11 & 343.22 & 333.44 & 351.53 & 332.11 & 333.00 & 351.07 & 334.88 & 353.02 \\
\ce{H2O}   & 146.62 & 165.08 & 150.73 & 169.19 & 148.86 & 150.50 & 168.96 & 153.41 & 172.05 \\
\ce{H2S}   & 147.84 & 165.88 & 153.49 & 171.55 & 152.77 & 153.57 & 171.61 & 154.91 & 173.04 \\
\ce{CO}    & 122.19 & 140.29 & 126.09 & 144.22 & 124.77 & 126.36 & 144.48 & 123.06 & 141.26 \\
\ce{N2}    & 99.75  & 117.93 & 102.99 & 121.23 & 101.74 & 102.71 & 120.95 & 100.96 & 119.32 \\
\ce{CO2}   & 108.47 & 126.89 & 113.32 & 131.81 & 112.47 & 112.52 & 130.99 & 115.65 & 134.34 \\
\ce{CH2O}  & 149.26 & 167.66 & 155.93 & 174.34 & 154.70 & 155.22 & 173.62 & 162.60 & 181.20 \\
\ce{2F-}   & 401.00 & 419.37 & 403.42 & 421.81 & 398.37 & 403.08 & 421.44 & 406.28 & 424.72 \\
\bottomrule
\end{tabular}
\caption{PAs of the studied molecules with the PW92, PBE0, B3LYP, and r$^2$SCAN functionals and HF, with and without epc17-2 electron-proton correlation. The data were obtained with the uncHq-aug-pc-4 electronic basis set and the PB6-H protonic basis set. All values are in kcal/mol.}
\label{tab:other-pas}
\end{table*}

\subsection{Other families of electronic basis sets \label{sec:otherfamilies}}

For this part of the work, we revert to using reference values obtained with the uncHq-aug-pc-4 electronic basis
set and the PB4-F1 protonic basis set, which we just showed to be at the CBS limit.
Moreover, as the results are the same across different types of functionals, we again restrict the discussion to the  electronic B3LYP functional and the epc17-2 electron-proton correlation functional.

\subsubsection{Correlation-consistent basis sets \label{sec:ccbasis}}

\begin{figure}
\begin{centering}
\includegraphics[width=\linewidth]{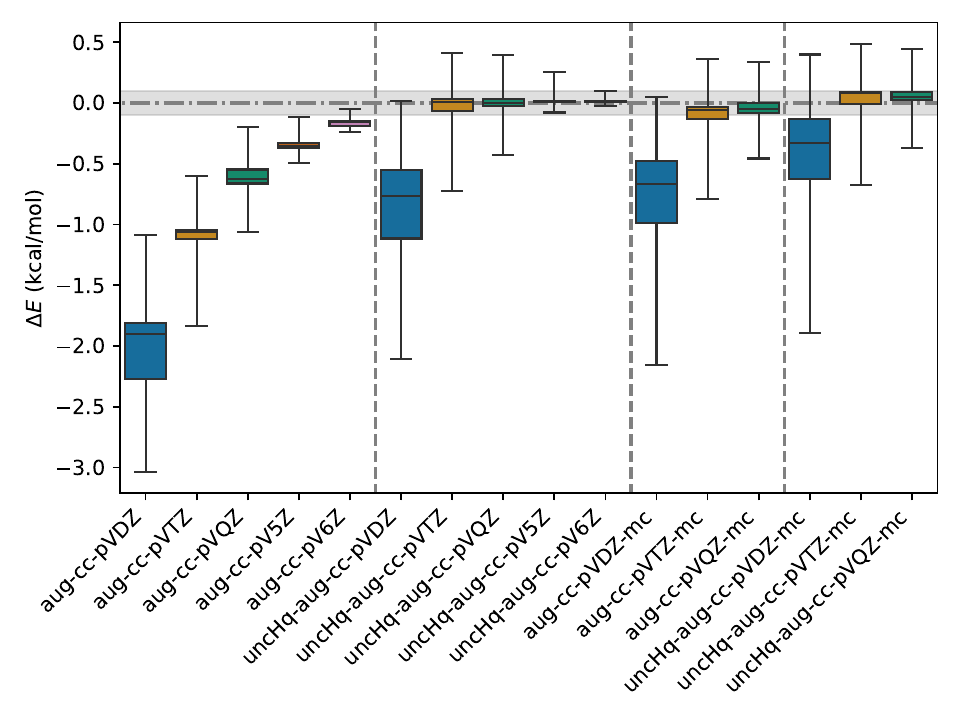}
\end{centering}
\caption{BSTEs of PAs obtained with the correlation-consistent basis set family and the PB4-F1 protonic basis set. All data are reported relative to uncHq-aug-pc-4/PB4-F1 values. The plot and box parameters are the same as in \cref{fig:pc}}
\label{fig:cc}
\end{figure}

Most of the calculations with the non-BO approach published so far appear to have employed the correlation-consistent basis set family.
Results of calculations with these basis sets are shown in \cref{fig:cc}.
Uncontracting the electronic basis set on the quantum hydrogens again results in marked reductions in the basis set truncation error.
A similar trend is once again observed where the uncontracted basis set outperforms the contracted basis of
a higher $\zeta$-level.
Importantly, the data for the largest basis set (uncHq-aug-cc-pV6Z) agree with the uncHq-aug-pc-4 reference values to within 0.10 kcal/mol, our stated aim in precision, again showing that we have reached the CBS limit.

\subsubsection{Multicomponent augmentations to correlation-consistent basis sets \label{sec:multicomponent}}

We now turn to the proposed augmentations of \citeit{Samsonova2023_AO_5033} aimed at improving the precision of results of non-BO-DFT calculations by adding functions to the electronic basis set to describe the proton density.
The results obtained using these basis sets are shown in \cref{fig:cc}.
As the multicomponent basis sets have a similar level of precision in either contracted or uncontracted form, uncontracting these augmented basis sets clearly has a smaller effect than those observed for the unmodified electronic basis sets in \cref{sec:electroniclimit,sec:ccbasis}, proving that the additional functions are indeed doing something.

However, as comparison to the results in \cref{fig:cc} shows, largely the same level of precision can be reached by simply uncontracting the original Dunning electronic basis set on the quantum hydrogen.
For example, the maximum basis set truncation error for the original aug-cc-pVQZ basis set is 1.06 kcal/mol, while the analogous value for aug-cc-pVQZ-mc is 0.46 kcal/mol which further reduces to 0.44 kcal/mol by uncontraction.
However, the basis set truncation error of uncHq-aug-cc-pVQZ is 0.43 kcal/mol, showing that the mc augmentation functions are unnecessary for these calculations.
Moreover, uncontraction of aug-cc-pVQZ only adds two s exponents (two basis functions), whereas the mc modification to aug-cc-pVQZ adds nine: three s functions, two p, two d, and one f function, resulting in a total of 29 additional basis functions.
It thus appears that standard electronic basis sets are suitable for multicomponent PA calculations, as long as one uncontracts them when necessary.

\subsubsection{Mixed basis sets \label{sec:mixedbasis}}
As already mentioned in \cref{sec:Introduction}, a common technique in non-BO calculations in the literature is to use a larger basis set for the quantum protons than for the rest of the molecule.\cite{Yang2018_JPCL_1765}
However, \citeit{Fowler2022_JCTC_7298} showed that the use of mixed basis sets leads to misleading proton affinities in coupled-cluster calculations.
We confirm this finding in our non-BO-DFT calculations below.

\begin{figure}
\begin{centering}
\includegraphics[width=\linewidth]{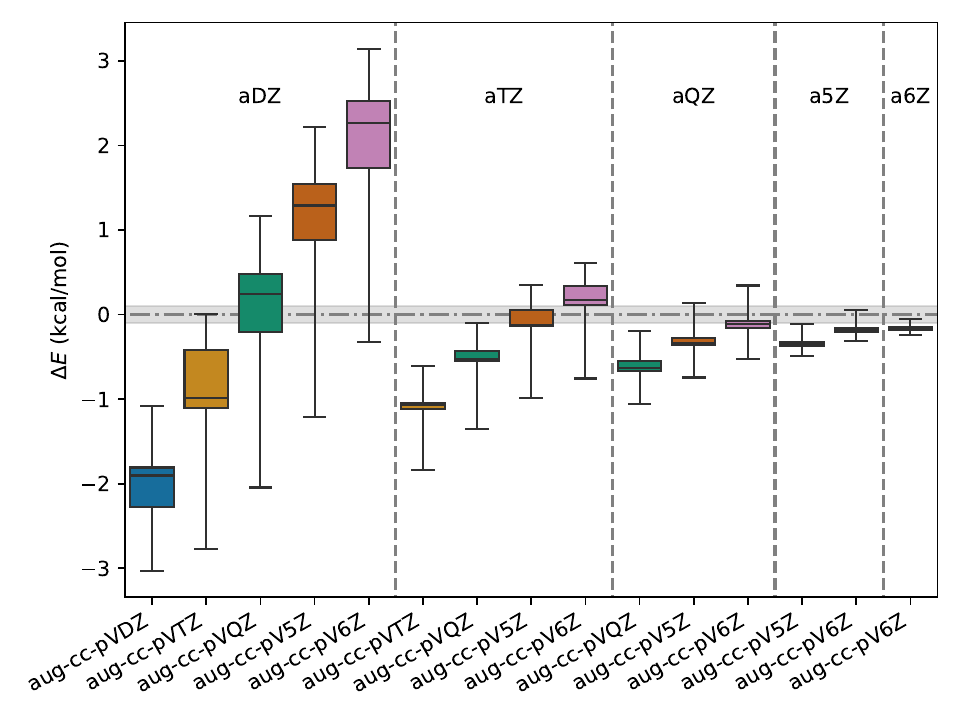}
\end{centering}
\caption{BSTEs of PAs obtained with mixed basis set where a larger correlation-consistent basis set is used on the quantum proton than that used on the other atoms. The floating text shows the basis set on the other atoms (aug-cc-pVXZ abbreviated as aXZ); the full aug-cc-pV6Z result is replicated only for comparison. The plot and box parameters are the same as in \cref{fig:pc}}
\label{fig:mixed}
\end{figure}

\Cref{fig:mixed} shows the BSTEs of PAs of molecules with a mixed basis set.
We see that the benefit is merely coincidental, exemplified by the rising trend in the case where the system has a double-$\zeta$ basis set and the quantum proton has an increasingly larger basis set.
The proton affinities are too small with a consistent choice for the contracted electronic basis set on all atoms, whereas enriching the basis set on the quantum proton leads to a systematic positive bias.
The total error can thereby be reduced by counterbalancing these two biases, but the minimum error corresponds to a Pauling point.\cite{Loewdin1986_IJQC_19}

\begin{figure}
\begin{centering}
\includegraphics[width=\linewidth]{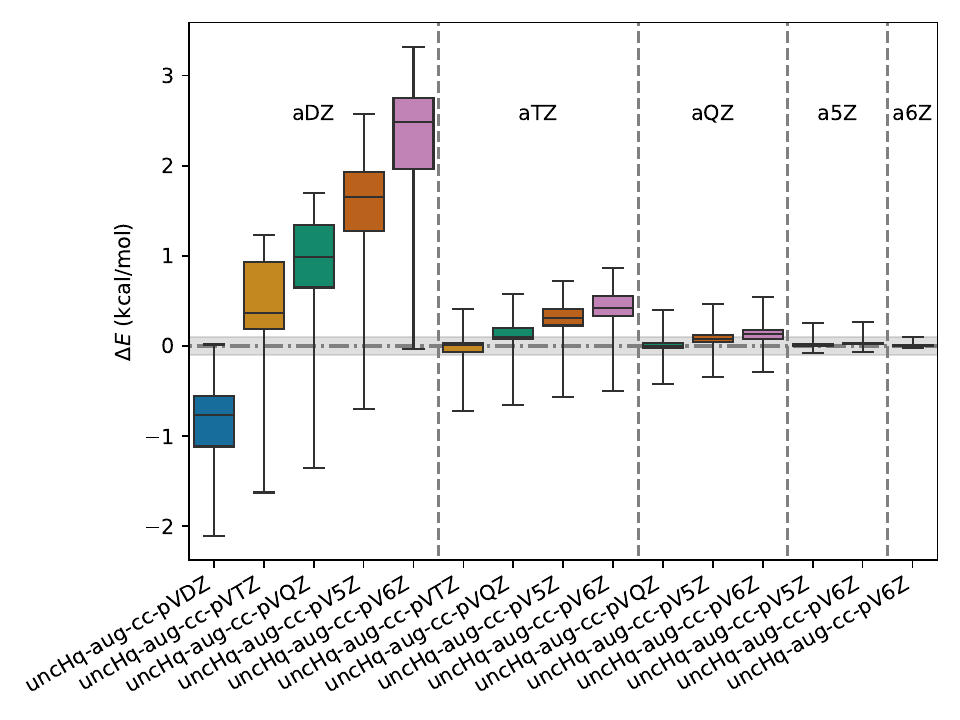}
\end{centering}
\caption{BSTEs of PAs obtained with mixed basis set where a larger correlation-consistent basis set is used on the quantum proton than that used on the other atoms, when the basis set on the quantum proton is also uncontracted. The floating text shows the basis set on the other atoms (aug-cc-pVXZ abbreviated as aXZ); the full uncHq-aug-cc-pV6Z result is replicated only for comparison. The plot and box parameters are the same as in \cref{fig:pc}}
\label{fig:mixed_unc}
\end{figure}

This is obvious from the uncontracted results shown in \ref{fig:mixed_unc}. As the proton is now more tightly bound thanks to more flexibility in the electronic wave function near the center of the delocalized proton, the negative bias is eliminated and there is no benefit to using a mixed basis set anymore.

\subsubsection{Karlsruhe basis sets \label{sec:def2}}

\begin{figure}
\begin{centering}
\includegraphics[width=1\linewidth]{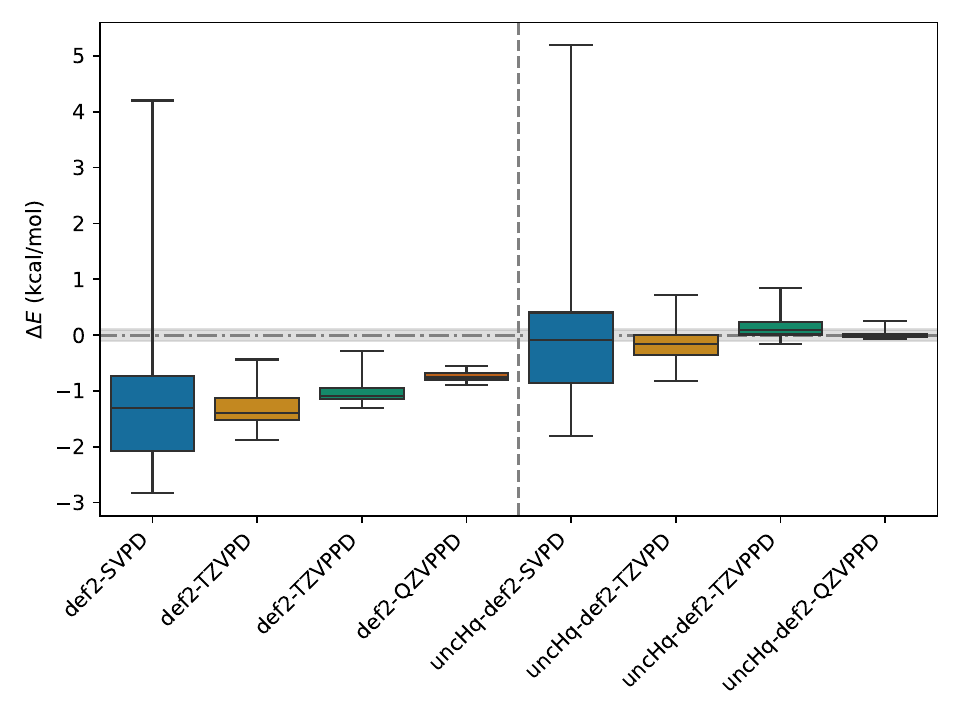}
\end{centering}
\caption{BSTEs of PAs obtained with the Karlsruhe electronic basis sets and the PB4-F1 protonic basis set. All data are reported relative to uncHq-aug-pc-4/PB4-F1 values. The plot and box parameters are the same as in \cref{fig:pc}}
\label{fig:def2}
\end{figure}

The Karlsruhe def2 basis sets have also been employed in some non-BO studies.
Analogous data for these basis sets are shown in \cref{fig:def2}.
These results tell a similar story as those for the cc basis sets discussed above in \cref{sec:ccbasis}.
Noticeable reductions in the basis set truncation error are once again observed when uncontracting the basis set on the quantum protons.
The uncHq-def2-SVPD basis set stands out with a considerably larger error than in the uncHq-aug-cc-pVDZ basis set.
An in-depth analysis shows that this error is due to a single outlier (HCN).
The error distributions for the triple-$\zeta$ and quadruple-$\zeta$ basis sets are closer in line with their correlation-consistent counterparts.

The basis set truncation errors for the largest basis set (uncHq-def2-QZVPPD) can be as large as 0.26 kcal/mol, which is similar to the precision of the aug-cc-pV6Z and uncHq-aug-cc-pV5Z basis sets.

\section{Summary and Discussion \label{sec:summary}}

We studied reaching the basis set limit in non-BO density functional theory (DFT) calculations of proton affinities (PAs).
Non-BO-DFT calculations require two types of basis sets to be employed: electronic and protonic.
We considered a variety of electronic basis sets: the Jensen polarization-consistent basis sets,\cite{Jensen2001_JCP_9113} Dunning's correlation consistent basis sets,\cite{Dunning1989_JCP_1007} Karlsruhe basis sets,\cite{Weigend2005_PCCP_305} as well as mixed basis sets where the quantum proton has a larger basis set than the rest of the molecule.
Diffuse electronic basis functions were included on all atoms for each family considered.
We also examined a variety of protonic basis sets: the protonic basis (PB) sets of \citeit{Yu2020_JCP_244123} and the various even-tempered basis sets of \citeitcomma{Yang2017_JCP_114113} and \citeitperiod{Khan2025_JCC_70082}

As the quantum protons no longer behave like point particles and are instead delocalized over a finite spatial domain, the electrons behave differently close to the center of the nuclear charge distribution.
The literature of nuclear spin-spin-coupling\cite{Helgaker1998_TCA_175} and x-ray calculations\cite{Besley2009_JCP_124308} as well as previous exploratory findings\cite{Nakai2003_JCP_1119, Nakai2005_JCP_164101, Moncada2025_JCP_24110} suggest a straightforward improvement of the electronic basis in non-BO calculations simply by uncontracting the basis functions centered on the quantum protons.
We found this to be emphatically the case, and all tested basis sets followed the same trend.
Uncontracting the electronic basis set on the quantum hydrogens significantly decreases the basis set truncation error, and allows obtaining results of at least one $\zeta$-level higher precision with negligible additional cost.
Although special augmentations of the electronic basis set on hydrogen have been proposed for non-BO calculations,\cite{Samsonova2023_AO_5033} we found uncontracting the standard basis to add considerably fewer basis functions, while yielding more precise results.

Mixing basis sets is not a generally accepted strategy in quantum chemistry, since using unbalanced combination of basis sets can lead to systematic errors.
\Cref{fig:mixed} showed that the improvement achieved by using a larger electronic basis set on the quantum proton was caused by the elimination of systematic error arising from unsatisfactory portrayal of tight functions, thus corresponding to a Pauling point.
This error is fully eliminated by uncontracting the basis on the quantum hydrogens, and consistent pairings of uncontracted basis sets lead to the lowest errors relative to the complete basis set (CBS) limit.

We were able to converge the non-BO-DFT BSTEs of PAs to 0.1 kcal/mol with respect to the basis set limit with two different families of electronic basis sets, as well as two different families of protonic basis sets.
We also showed that changing one basis set while keeping the other fixed does not change the obtained PAs, as long as large enough electronic and protonic basis sets are employed, proving that we reached the CBS limit.

Our results point out deficiencies in existing approaches and guide the way for the design of better computational procedures.
We find that the protonic basis sets of \citeit{Yu2020_JCP_244123} do not represent a systematically convergent hierarchy in proton affinity, and that the PB5 basis sets appear to yield results of poorer quality than those obtained with the PB4 or PB6 basis sets.
We tentatively attributed these issues to the use of inconsistent methodologies to optimize these basis sets, which likely correspond to Pauling points.

We find that more work is needed to determine cohesive and systematic protonic basis sets.
Clear error-balanced pairings with electronic basis sets are needed for an optimal approach to non-Born--Oppenheimer calculations.
The basis sets of \citeit{Yu2020_JCP_244123} appear to be too large for double-$\zeta$ or triple-$\zeta$ electronic basis sets.
Based on the data in \cref{fig:PB,fig:epc}, it appears that while p and d functions are clearly required for high precision, protonic f functions do not appear to be important for the presently studied ground-state non-BO-DFT calculations.

Following the established principles of basis set design,\cite{Dunning1989_JCP_1007, Jensen2001_JCP_9113} the electronic and protonic basis sets should be chosen in a way that leads to similar errors in the protonic and electronic parts of the wave function.
Comparing the errors made in the electronic discretization in \cref{fig:pc} and the protonic discretization in \cref{fig:PB,fig:epc} suggest  the following pairings of the protonic basis set to an electronic basis set of given quality: an electronic polarized double-$\zeta$ basis should likely employ an s-function only protonic basis set, while the worst-case precision of an electronic polarized triple-$\zeta$ basis set appears to match that of a protonic sp basis set.
An electronic polarized quadruple-$\zeta$ basis set appears to match the precision of a protonic spd basis set.
It also appears that there are no protonic basis sets suitable for higher electronic $\zeta$-levels; we hope to address this deficiency in future work.
We also hope to report optimally balanced electronic and protonic basis sets in future work.
While this work was focused on proton affinities with non-BO-DFT, we also plan to investigate the basis set convergence of other properties and other levels of theory in the future.

\section*{Supporting Information Available}
The employed optimized geometries in \AA{}ngstr\"om for the molecules, as well as the calculated Born--Oppenheimer and non-Born--Oppenheimer total energies in Hartrees are available in a file in the supporting information in JavaScript Object Notation (JSON) format for each of the studied basis set combinations.
All the results of this work are reproducible from these data.

\section*{Data Availability Statement}
The data that support the findings of this study are available within the supplementary material.

\section*{Acknowledgments}
We thank Fabijan Pavo\v{s}evi{\'c}---who was involved in this project at an early stage---for invaluable discussions on non-BO calculations in general, and the present manuscript in specific.
SL thanks Roland Lindh for discussions on Pauling points.
We thank the Research Council of Finland for financial support through the Finnish Quantum Flagship, project number 358878, as well as through the academy fellowship of SL, project numbers 350282, 353749, and 368487.
We thank CSC---IT Center for Science Ltd (Espoo, Finland) for computational resources.

\bibliography{citations,extras}

\end{document}